\documentclass{article}

\usepackage{PRIMEarxiv}

\usepackage[utf8]{inputenc} % allow utf-8 input
\usepackage[T1]{fontenc}    % use 8-bit T1 fonts
\usepackage{url}            % simple URL typesetting
\usepackage{booktabs}       % professional-quality tables
\usepackage{amsfonts}       % blackboard math symbols
\usepackage{nicefrac}       % compact symbols for 1/2, etc.
\usepackage{microtype}      % microtypography
\usepackage{lipsum}
\usepackage{fancyhdr}       % header
\usepackage{graphicx}       % graphics
\graphicspath{{media/}}     % organize your images and other figures under media/ folder

\usepackage{amsmath}
\usepackage{array}
\usepackage{booktabs}
\usepackage{multirow}
\usepackage{amssymb}
\usepackage{bm}
\usepackage{textcomp}
\usepackage{float}
\usepackage{algorithm}
\usepackage{algpseudocode}
\usepackage{makecell}

\usepackage{times}  % DO NOT CHANGE THIS
\usepackage{helvet}  % DO NOT CHANGE THIS
\usepackage{courier}  % DO NOT CHANGE THIS
\usepackage{natbib}  % DO NOT CHANGE THIS AND DO NOT ADD ANY OPTIONS TO IT
\usepackage{caption}
\usepackage[colorlinks=true, linkcolor=black, citecolor=blue, urlcolor=blue]{hyperref}

\title{BConformeR: A Conformer Based on Mutual Sampling for Unified Prediction of Continuous and Discontinuous Antibody Binding Sites
}

\author{
  Zhangyu You\thanks{%\textit{*}: 
  \textbf{Equal contribution.}}\,\,, Jiahao Ma$^*$,\\
  Material Innovation Institute for Life Sciences and Energy\\
  The University of Hong Kong\\
  Hetao SZ-HK Cooperation Zone\\
  \texttt{yzy48940@gmail.com}, \texttt{jiahao.ma@connect.hku.hk} \\
  \And
  Hongzong Li$^*$, \\
  Hong Kong Generative AI Research and Development Center \\
  The Hong Kong University of Science and Technology \\
  Hong Kong\\
  \texttt{lihongzong@ust.hk} \\ 
  \And
  Ye-Fan Hu$^*{^,}$\thanks{%\textit{\dagger}: 
  \textbf{Corresponding authors.}}\,\,,\\
  Computational Immunology Centre \\
  BayVax Biotech Limited \\
  Hong Kong\\
  \texttt{yefan.hu@bayvaxbio.com} \\  
  \And
  Jian-Dong Huang$^\dagger$\\
  School of Biomedical Sciences \\
  The University of Hong Kong \\
  Hong Kong\\
  \texttt{jdhuang@hku.hk} \\
}

\begin{document}
\maketitle
\begin{abstract}
Accurate prediction of antibody-binding sites (epitopes) on antigens is crucial for vaccine design, immunodiagnostics, therapeutic antibody development, antibody engineering, research into autoimmune and allergic diseases, and advancing our understanding of immune responses. Despite in silico methods that have been proposed to predict both linear (continuous) and conformational (discontinuous) epitopes, they consistently underperform in predicting conformational epitopes. In this work, we propose Conformer-based models trained separately on AlphaFold-predicted structures and experimentally determined structures, leveraging convolutional neural networks (CNNs) to extract local features and Transformers to capture long-range dependencies within antigen sequences. Ablation studies demonstrate that CNN enhances the prediction of linear epitopes, and the Transformer module improves the prediction of conformational epitopes. Experimental results show that our model outperforms existing baselines in terms of MCC, ROC-AUC, PR-AUC, and F1 scores on both linear and conformational epitopes. Code is available at \href{https://anonymous.4open.science/r/bconformer-epitope-519D}{https://anonymous.4open.science/r/bconformer-epitope-519D}.
\end{abstract}


% %------------------------------------------------------------------------

\section{Introduction}

B cells are key effectors of the adaptive immune system, coordinating responses to pathogenic and aberrant signals. Each B cell expresses a single clonotype of B-cell receptor (BCR), a membrane-bound antibody that interacts with specific antigens. The binding sites on the antigen—known as the epitope—can be linear (continuous) or conformational (discontinuous), and the corresponding binding sites on the antibody are called paratopes. Linear epitopes are formed by linear stretches of amino acid residues, while conformational epitopes are formed by distant residues brought together in 3D through folding.

% %------------------------------------------------------------------------
\begin{figure}[t]
\centering
\includegraphics[width=0.65\textwidth]{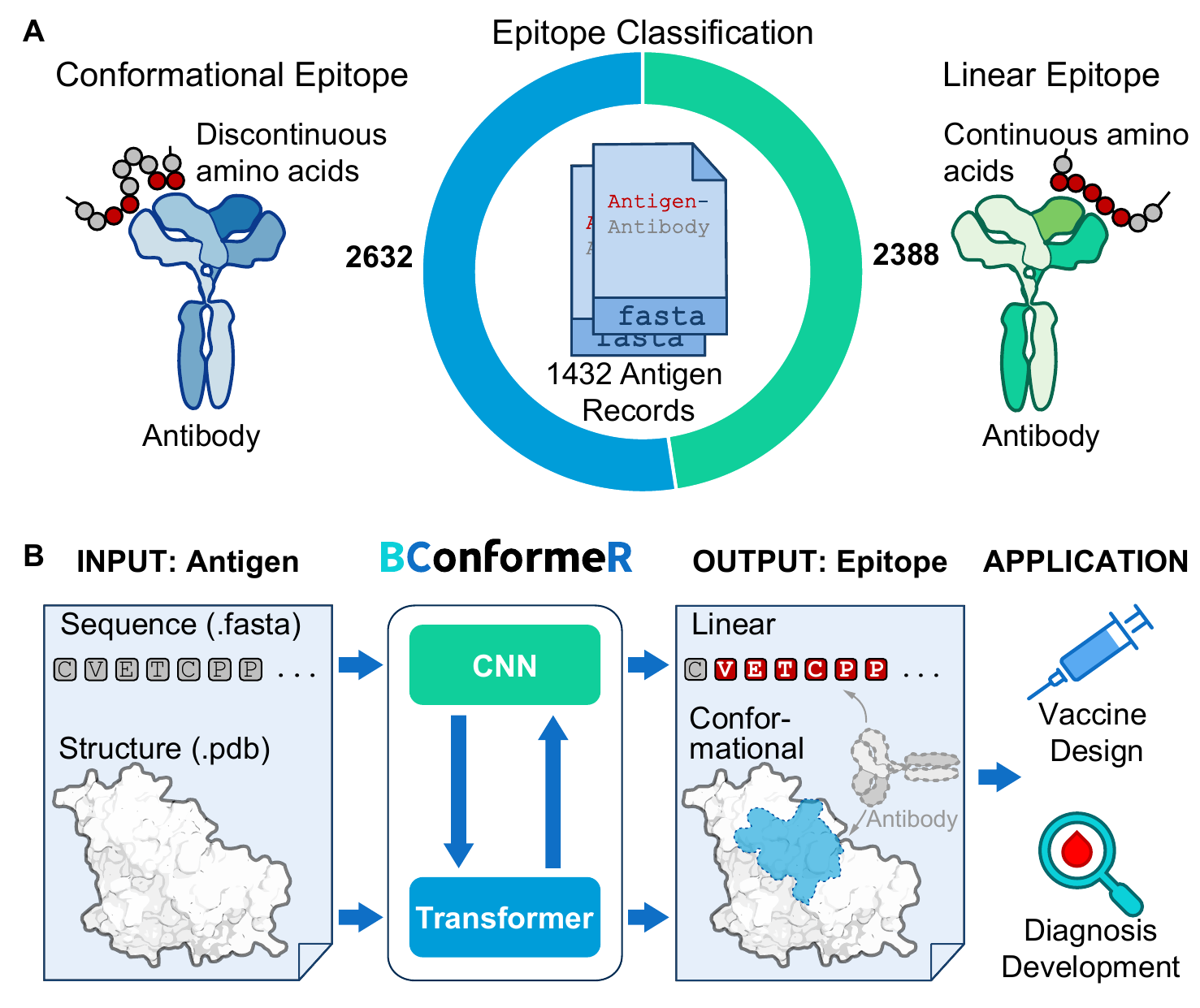} 
\caption{Overview of epitope classification and BConformeR prediction.
(A) Epitopes, i.e., Antibody-binding sites, include conformational fragments (n=2,632) and linear regions (n=2,388). Data were derived from 1,432 antigen records.
(B) Epitope prediction workflow. Antigen inputs are processed to predict linear and conformational epitopes, which enables applications in vaccine design and diagnostic development.
}
\label{fig1}
\end{figure}

% %------------------------------------------------------------------------

Identifying epitopes is critical for antibody engineering \cite{koz:18,ji:22}, disease diagnosis \cite{va:22,ta:24}, and vaccine optimization \cite{dv:24,ba:25}. Experimental techniques such as X-ray crystallography and cryo-electron microscopy provide high-resolution antigen-antibody (Ag–Ab) interactions but are resource-intensive \cite{crys:21,cryoEM:25}. Phage display is faster but lacks atomic-level precision \cite{pd:22}. Therefore, many in silico methods have been developed to predict epitopes.

Sequence-based methods such as SEMA-1D~\cite{SEMA,SEMA2.0}, BepiPred~\cite{BepiPred1.0,BepiPred2.0,BepiPred3.0} and SEPPA~\cite{SEPPA3.0} are widely used due to the availability of antigen sequences, with SEMA-1D 2.0~\cite{SEMA2.0} and BepiPred-3.0~\cite{BepiPred3.0} notably improving performance by incorporating representations from the protein language model ESM-2~\cite{ESM-2pre,ESM-2}. Structure-based methods, such as SEMA-3D~\cite{SEMA,SEMA2.0}, DiscoTope~\cite{DiscoTope2.0,DiscoTope3.0}, ScanNet~\cite{ScanNet} and Epitope3D~\cite{Epitope3D}, take the residue geometry and surface accessibility into consideration, but remain constrained by the limited availability of high-quality antigen structures. Among these methods, DiscoTope-3.0~\cite{DiscoTope3.0} achieves the best performance in the benchmark evaluation.

% %------------------------------------------------------------------------
\begin{figure*}[t]
\centering
\includegraphics[width=1\textwidth]{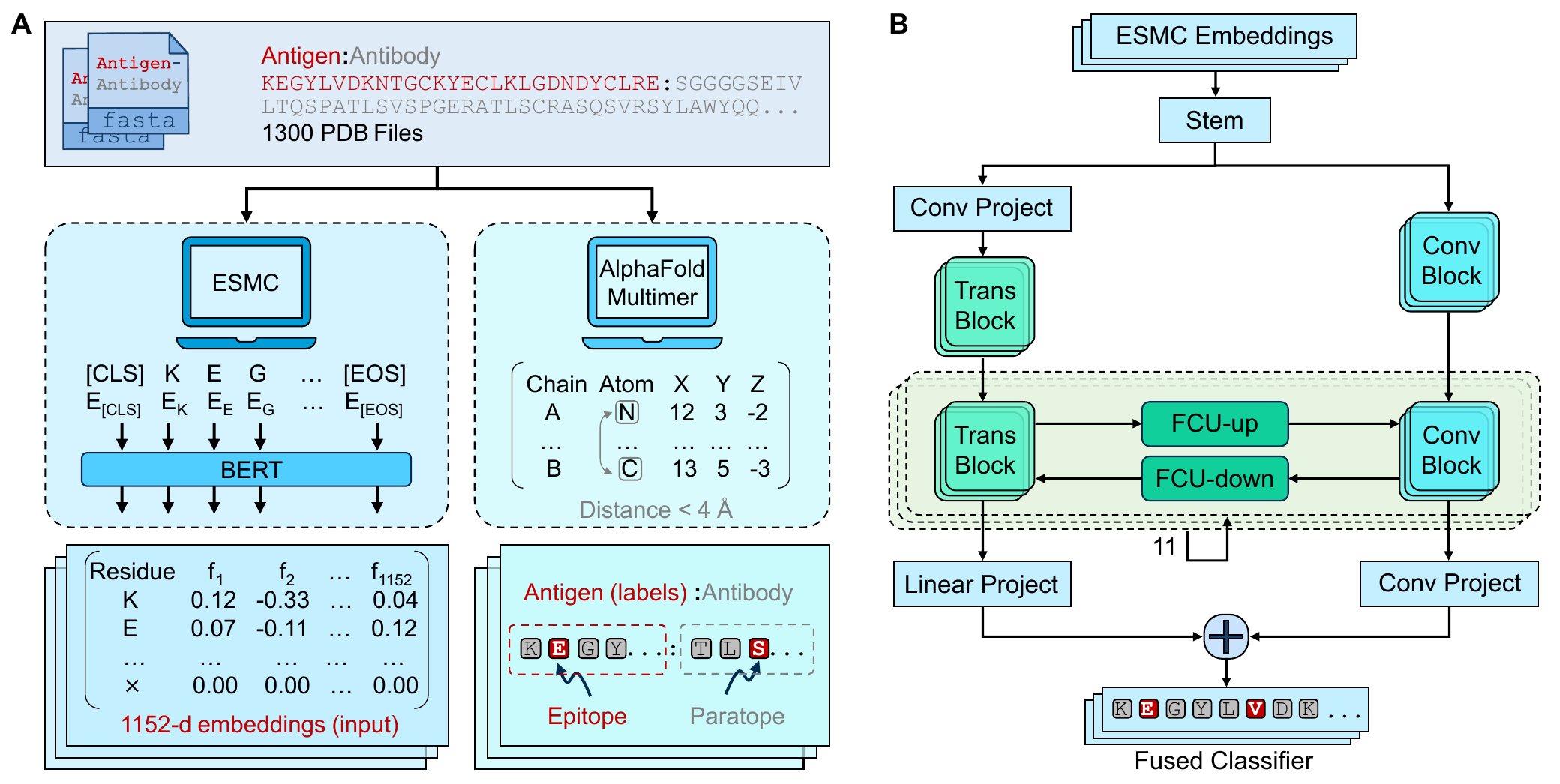} % Reduce the figure size so that it is slightly narrower than the column.
\caption{BConformeR overview. (A) Preparation pipeline. (Right) Preparation of Alphafold-predicted structures and epitope extraction. (Left) Antigen embeddings. (B) Architecture overview. Stem module applies convolution and max-pooling to the ESM Cambrian (ESMC) embeddings; The Conv block and Trans block are based on ResNet bottleneck and Vision Transformer modules respectively; FCU-Up and FCU-Down connect the Conv and Trans blocks; The classifier linearly fuses features from both branches.}
\label{fig2}
\end{figure*}

% %------------------------------------------------------------------------

However, epitope prediction remains a challenging task, with existing methods underperform on conformational ones. Moreover, there is a lack of hybrid backbone models to effectively handle both linear and conformational epitopes. A key source of this complexity stems from the nature of Ag-Ab interfaces, where linear epitopes are clustered along the sequence, and conformational epitopes are often spatially proximate but dispersed across different regions of the linear sequence \cite{OverviewEp}. To address this multi-scale complexity, we designed the Conformer architecture, which integrates convolution layers to learn local residue-level features and self-attention mechanisms to capture long-range dependencies within antigen sequences \cite{Conformer20,Conformer21}.

Here, we propose \textbf{BConformeR}, a \textbf{Conformer}-based model for continuous and discontinuous \textbf{BCR} epitope prediction. BConformeR outperforms baselines—SEPPA-3.0, BepiPred-3.0, SEMA-1D 2.0 and DiscoTope-3.0—across multiple metrics, including MCC, ROC-AUC, PR-AUC, and F1 score on both linear and conformational epitopes.
To further dissect the model’s design, we ablated ResNet bottlenecks and ViT modules from Conformer, showing that convolutional layers improve linear epitope prediction, and Transformer modules are critical for capturing discontinuous epitope patterns.

% %------------------------------------------------------------------------
% However, existing models rarely incorporate both local features and long-range dependencies within antigen sequences, showing limited performance on conformational epitopes. Therefore, we adopted a Conformer based on mutual sampling, employing CNNs to extract local features and Transformers to capture long-range dependencies within antigen sequences.

% \subsection{Antibody-aware evaluation models} Despite advances in epitope prediction, evaluating Ag-Ab binding confidence remains challenging, which is crucial for vaccine selection. AbEpitope-1.0, a recent method for predicting antibody targets by use of AlphaFold and inverse folding, significantly improving the prediction accuracy of Ag-Ab interfaces. They constructed negative Ag-Ab samples (false binding pairs) by swapping antibody chains across real complexes in the dataset. Each sample was labeled as either true or false to train the model to distinguish authentic Ag-Ab interactions from decoys. The final model, AbEpiTarget-1.0, achieved superior performance in antibody-sensitive scoring for target prediction \cite{AbEpitope}. In this work, we utilize the Ag–Ab complex dataset provided by AbEpitope-1.0 to construct an AlphaFold-predicted structure dataset (Fig.~\ref{fig2}A).

% %------------------------------------------------------------------------

\section{Methods}
\subsection{Epitope classification}
To extract epitope residues, we define the residue–neighbor, where an antigen residue and an antibody residue are considered binding sites if the minimum distance between any pair of their heavy atoms is less than 4 Å \cite{epitope_id}. Generally, epitopes are separated into multiple Regions (R) and Fragments (F) for each antigen sequence. Regions are defined as linear stretches of antigen sequence having at least three antibody-contacting residues. Gaps between contacting residues are allowed, and a gap size of up to three non-contacting residues is chosen on the basis of the structure of $\alpha$-helices, which enables residues on the same helical face to be included \cite{epitope_cl}. Fragments, which do not contain enough antibody-contacting residues to qualify as a region, are considered discontinuous in our study (Fig.~\ref{fig1}A).

For real and predicted annotations, linear and discontinuous epitopes are extracted from regions and fragments, respectively. Ambiguous cases occur when the real and predicted labels differ, especially when a residue is annotated as part of a region in the ground truth but as part of a fragment in the prediction, or vice versa. To address this, all classifications are determined based on the ground-truth labels. That is, for a real label \( r \in \{N, L, D\} \) and predicted label \( p \in \{N, L, D\} \), the final classification \(c(r, p)\) is defined as:
\begin{equation}
c(r, p) =
\begin{cases}
L, & (r, p) \in \{(L, \cdot), (N, L)\} \\
D, & (r, p) \in \{(D, \cdot), (N, D)\} \\
N, & (r, p) = (N, N)
\end{cases}
\label{epiclass}
\end{equation}
where \(N\), \(L\), and \(D\) denote non-epitope, linear epitope and discontinuous epitope, respectively. Performance metrics (\textit{e.g.}, F1) are then calculated for each type of epitopes.

% % %------------------------------------------------------------------------

\renewcommand{\arraystretch}{1.5}
\setlength{\tabcolsep}{4pt}

\begin{table}[!t]
\centering
\makebox[\textwidth][c]{%
\resizebox{0.9\textwidth}{!}{%
\begin{tabular}{c|c|c|c|c|c|c|c}
\Xhline{1pt}
 & \multicolumn{3}{c|}{\textbf{stage 1}} & \multicolumn{2}{c|}{\textbf{stage 2}} & \multicolumn{2}{c}{\textbf{stage 3}} \\
\Xhline{0.8pt}

\textbf{CNN} &
\raisebox{0.21em}{\footnotesize$
\left[
\begin{array}{l}
1,144\\[-.2em]
3,144\\[-.2em]
1,576
\end{array}\right]$} &
\raisebox{0.21em}{\footnotesize$
\left[
\begin{array}{l}
1,144\\[-.2em]
3,144\\[-.2em]
1,576
\end{array}\right]$} &
\raisebox{0.21em}{\footnotesize$
\left[
\begin{array}{l}
1,144\\[-.2em]
3,144\\[-.2em]
1,576
\end{array}\right]$} &

\raisebox{0.21em}{\footnotesize$
\left[
\begin{array}{l}
1,288\\[-.2em]
3,288\\[-.2em]
1,1152
\end{array}\right]$} &
\raisebox{0.21em}{\footnotesize$
\left[
\begin{array}{l}
1,288\\[-.2em]
3,288\\[-.2em]
1,1152
\end{array}\right]$} &
\raisebox{0.21em}{\footnotesize$
\left[
\begin{array}{l}
1,576\\[-.2em]
3,576\\[-.2em]
1,2304
\end{array}\right]$} &
\raisebox{0.21em}{\footnotesize$
\left[
\begin{array}{l}
1,576\\[-.2em]
3,576\\[-.2em]
1,2304
\end{array}\right]$} \\
\Xhline{0.8pt}

\textbf{FCU} &
$-$ &
\multicolumn{1}{c}{\raisebox{0.3em}{\footnotesize$
\begin{array}{c}
[1,1536]\\[-.2em]
\big\downarrow
\end{array}$}} &
\raisebox{0.3em}{\footnotesize$
\begin{array}{c}
\big\uparrow\\[-.2em]
[1,144]
\end{array}$} &
\multicolumn{1}{c}{\raisebox{0.3em}{\footnotesize$
\begin{array}{c}
[1,1536]\\[-.2em]
\big\downarrow
\end{array}$}} &
\raisebox{0.3em}{\footnotesize$
\begin{array}{c}
\big\uparrow\\[-.2em]
[1,288]
\end{array}$} &
\multicolumn{1}{c}{\raisebox{0.3em}{\footnotesize$
\begin{array}{c}
[1,1536]\\[-.2em]
\big\downarrow
\end{array}$}} &
\raisebox{0.3em}{\footnotesize$
\begin{array}{c}
\big\uparrow\\[-.2em]
[1,576]
\end{array}$} \\
\Xhline{0.8pt}

\textbf{Trans} &
\raisebox{0.21em}{\footnotesize$
\left[
\begin{array}{c}
\text{MHSA-24, 1536}\\[-.2em]
1,3072\\[-.2em]
1,1536
\end{array}\right]$} &
\multicolumn{2}{c|}{\raisebox{0.21em}{\footnotesize$
\left[
\begin{array}{c}
\text{MHSA-24, 1536}\\[-.2em]
1,3072\\[-.2em]
1,1536
\end{array}\right]$}} &
\multicolumn{2}{c|}{\raisebox{0.21em}{\footnotesize$
\left[
\begin{array}{c}
\text{MHSA-24, 1536}\\[-.2em]
1,3072\\[-.2em]
1,1536
\end{array}\right]$}} &
\multicolumn{2}{c}{\raisebox{0.21em}{\footnotesize$
\left[
\begin{array}{c}
\text{MHSA-24, 1536}\\[-.2em]
1,3072\\[-.2em]
1,1536
\end{array}\right]$}} \\
\Xhline{0.8pt}

\multicolumn{1}{c|}{} &
\multicolumn{1}{c|}{\(\times1\)} &
\multicolumn{2}{c|}{\(\times3\)} &
\multicolumn{2}{c|}{\(\times4\)} &
\multicolumn{2}{c}{\(\times4\)} \\
\Xhline{1pt}
\addlinespace[2mm]
\end{tabular}%
}}
\caption{Interaction layers of BConformeR. MHSA denotes multi-head self-attention.}
\label{table1}
\end{table}

% --------------------------------------------------

\subsection{Conformer}

\subsubsection{Overview} To take advantage of both local and global representations for epitope prediction, we construct BConformeR, a dual-branch architecture adapted from Conformer \cite{Conformer21}. BConformeR consists of a shared convolutional stem, a CNN branch, a Transformer branch, Feature Coupling Units (FCUs) for bidirectional feature exchange, and two per-position classifiers (Fig.~\ref{fig2}B).

Sequences are padded to match the longest length in each batch. The stem consists of a 1D convolution and a 1D max pooling, with the embedding dimension 
$C$ (\textit{e.g.}, 1152 generated by ESMC-600M) reduced to $C/4$ channels after the stem. Then as shown in Table~\ref{table1}, both branches process initial features and no feature coupling is applied in the first block. From the second block onward, FCUs are inserted to progressively fuse local and global representations. At each stage, CNN-derived features are passed to the Transformer branch, and Transformer-derived global representations are fed back to the CNN branch. The CNN branch gradually expands to $2C$ channels, and the Transformer branch maintains $D$-dimensional (\textit{e.g.}, 1536) token embeddings at the end.

Finally, a 1D convolutional classifier and a linear classifier are applied to the outputs of the CNN and Transformer branches, respectively, producing per-residue logits $\mathbf{y}_{\text{CNN}}, \mathbf{y}_{\text{Trans}} \in \mathbb{R}^{L \times 2}$. These are linearly fused and converted to per-residue class probabilities using the softmax function:
\begin{equation}
\mathbf{p}_{i} = \mathrm{softmax} \left( \alpha \mathbf{y}_{\text{CNN},i} + (1 - \alpha) \mathbf{y}_{\text{Trans},i} \right),
\label{fusion_softmax}
\end{equation}
where $\mathbf{y}_{i} \in \mathbb{R}^2$ denotes the logits for the $i$-th residue, and $\alpha$ is the weighted parameter.

\subsubsection{CNN branch} The CNN branch is composed of three stages after the stem, each consisting of repeated convolutional blocks based on the ResNet bottleneck design \cite{ResNet}. As shown in Table~\ref{table1}, given an input $\mathbf{x} \in \mathbb{R}^{C' \times L}$, the bottleneck performs three sequential convolutions with a residual connection applied to $\mathbf{x}$:
\begin{equation}
\begin{aligned}
\mathbf{z}_1 &= \sigma\left(\text{BN}_1\left(\text{Conv}_{1 \times 1}(\mathbf{x})\right)\right), \\
\mathbf{z}_2 &= \sigma\left(\text{BN}_2\left(\text{Conv}_{3 \times 1}(\mathbf{z}_1)\right)\right), \\
\mathbf{z}_3 &= \text{BN}_3\left(\text{Conv}_{1 \times 1}(\mathbf{z}_2)\right), \\
\mathbf{y} &= \sigma\left( \mathbf{z}_3 + \mathcal{F}(\mathbf{x}) \right),
\end{aligned}
\label{cnn_block}
\end{equation}
where $\sigma(\cdot)$ denotes the ReLU activation, BN represents 1D batch normalization, and 
\[
\mathcal{F}(\mathbf{x}) =
\begin{cases}
\mathbf{x}. & \text{if shape matches} \\
\text{BN}_{\text{res}}\left(\text{Conv}_{1 \times 1}(\mathbf{x})\right). & \text{otherwise}
\end{cases}
\]

The channel dimension is first downsampled, then expanded by a factor of 4 within each bottleneck. The first convolution block contains a single bottleneck, while subsequent blocks include bottlenecks before and after the Transformer branch.

\subsubsection{Transformer branch} The Transformer branch processes embeddings using 12 repeated Transformer blocks, with each block following a structure similar to ViT~\cite{ViT}. As shown in Table~\ref{table1}, given the input $\mathbf{x} \in \mathbb{R}^{L \times D}$, a single Transformer block updates the representation as follows:
\begin{equation}
\begin{aligned}
\mathbf{z}_1 &= \mathbf{x} + \mathrm{DropPath}\left(\mathrm{MHSA}\left(\mathrm{LN}_1(\mathbf{x})\right)\right), \\
\mathbf{z}_2 &= \mathbf{z}_1 + \mathrm{DropPath}\left(\mathrm{MLP}\left(\mathrm{LN}_2(\mathbf{z}_1)\right)\right),
\end{aligned}
\label{transformer_block}
\end{equation}
where $\mathrm{MHSA}(\cdot)$ denotes the multi-head self-attention mechanism with $h$ (\textit{e.g.}, 24) heads \cite{Attention}, and $\mathrm{MLP}(\cdot)$ consists of a linear up-projection to $2D$ dimensions, a GELU activation, followed by a linear down-projection to $D$. The $\mathrm{DropPath}$ operator implements stochastic depth regularization.

Since the CNN branch already encodes positional information, no explicit positional embeddings are added. Also, we omit the use of $[\text{CLS}]$ tokens as our task is residue-level classification. This branch facilitate the modeling of long-range interactions between residues, preserving spatial information of discontiunous epitopes.

\subsubsection{Feature coupling units} 
To enable bidirectional interaction between local CNN features and global Transformer representations, we introduce feature coupling units (FCUs) from the second block. FCU-Down projects CNN features into the Transformer space and adds them to the Transformer input:
\begin{equation}
\begin{aligned}
\mathbf{F}_{\mathrm{CNN}}^\rightarrow &= \left(\mathrm{BN}\big(\mathrm{Conv}_{1 \times 1}(\mathbf{F}_{\mathrm{CNN}})\big)\right)^T, \\
\widetilde{\mathbf{F}}_{\mathrm{Trans}} &= \mathbf{F}_{\mathrm{Trans}} + \mathbf{F}_{\mathrm{CNN}}^\rightarrow.
\end{aligned}
\label{fcu_down}
\end{equation}

FCU-Up maps Transformer features back to the CNN space and merges them:
\begin{equation}
\begin{aligned}
\mathbf{F}_{\mathrm{Trans}}^\leftarrow &= \mathrm{BN}\big(\mathrm{Conv}_{1 \times 1}(\widetilde{\mathbf{F}}_{\mathrm{Trans}}^T)\big), \\
\widetilde{\mathbf{F}}_{\mathrm{CNN}} &= \mathbf{F}_{\mathrm{CNN}} + \mathbf{F}_{\mathrm{Trans}}^\leftarrow.
\end{aligned}
\label{fcu_up}
\end{equation}

% %------------------------------------------------------------------------

\subsection{Calibrated Score}

Scores of residues in each antigen are normalized using the following formula:
\begin{equation}
z_{i} = \frac{s_i - \mu_\text{antigen}}{\sigma_\text{antigen}},
\end{equation}
where \(s_i\) denotes the raw score of residue \(i\), and \(\mu_\text{antigen}\) and \(\sigma_\text{antigen}\) represent the mean and standard deviation of scores for that antigen, respectively. Both \(\mu_\text{antigen}\) and \(\sigma_\text{antigen}\) are estimated using two separate generalized additive models (GAMs)~\cite{GAM}, with one predicting the mean residue score from antigen length and the other predicting the standard deviation from the mean residue score~\cite{DiscoTope3.0}.

% %------------------------------------------------------------------------

\section{Experiments}
\subsection{Experiments Setup}
\subsubsection{Dataset} 

We trained two models separately using experimentally solved and AlphaFold-predicted structures.
The dataset of experimentally solved Ag–Ab complexes was obtained from AACDB~\cite{AACDB_data}. Redundant complexes were removed by retaining only one complex for each PDB ID, resulting in a curated dataset of 3,674 Ag-Ab complexes for training. For the AlphaFold-based dataset, we collected 1,300 Ag–Ab complex FASTA files from AbEpiTope-1.0~\cite{AbEpitope} and modeled each complex using AlphaFold-2.3 ColabFold~\cite{AFMultimer}. Thirty structural predictions were generated for each complex, producing a total of 39,000 structures. For each complex, we retained the top-ranked model and discarded those with a pLDDT score below 0.82. After extracting epitopes for each complex, we further excluded those with insufficient epitope residues. The final dataset contained 1,180 Ag–Ab complexes, of which 1,080 were used for training BConformeR and 100 for validation and ablation studies.

\begin{figure}[!t]
\centering
\includegraphics[width=0.85\textwidth]{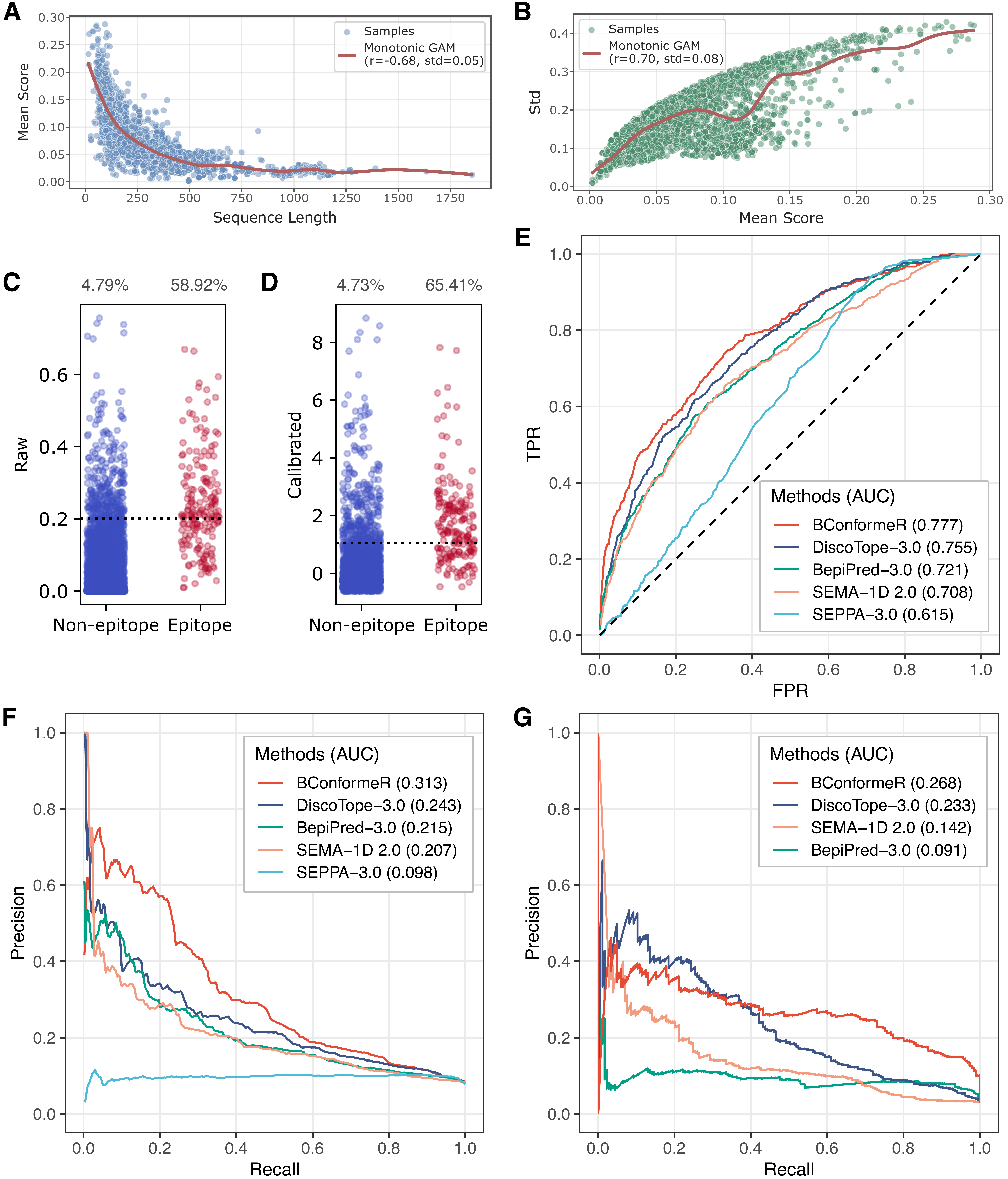} % Reduce the figure size so that it is slightly narrower than the column.
\caption{Improved performance of BConformeR.
(A–B) GAMs for estimating $\mu$ and $\text{std}$.  
(C–D) Comparison between raw and calibrated scores on the SARS-CoV-2  test set.
(E–F) ROC-AUC and PR-AUC; BConformeR trained on AlphaFold-predicted structures and evaluated on the blind test set of 24 antigens.  
(G) PR-AUC; BConformeR trained on experimentally solved structures and evaluated on the SARS-CoV-2 test set.
}
\label{fig3}
\end{figure}

As the datasets used among methods are not consistent, we assembled a SARS-CoV-2 test set from CoV-AbDab~\cite{CoV-AbDab} for models trained on experimentally solved structures, and selected 24 antigen records from the SEMA-1D 2.0 test sets to construct a blind test set for AlphaFold-based models.

% %------------------------------------------------------------------------

\subsubsection{Calibrated score} 

After training, the \(\mu_{\text{antigen}}\) and \(\sigma_{\text{antigen}}\) were estimated from the predicted score distributions (Fig.~\ref{fig3}A,B). The scores were then calibrated based on different antigen lengths, which effectively reduced false positives while improving the true positive rate (Fig.~\ref{fig3}C,D).

% %------------------------------------------------------------------------

\begin{table*}[!t]
\centering
\renewcommand{\arraystretch}{1}
\setlength{\tabcolsep}{4pt}
\resizebox{0.9\textwidth}{!}{
\begin{tabular}{ccccccc}
\toprule
\multirow{2}{*}{\textbf{Method}} & \multicolumn{3}{c}{\textbf{Alphafold structures}} & \multicolumn{3}{c}{\textbf{Solved structures}} \\
\cmidrule(lr){2-4} \cmidrule(lr){5-7}
 & \textbf{F1-L $\uparrow$} & \textbf{F1-D $\uparrow$} & \textbf{MCC $\uparrow$} & \textbf{F1-L $\uparrow$} & \textbf{F1-D $\uparrow$} & \textbf{MCC $\uparrow$} \\
\midrule
SEPPA-3.0 & 0.158 ± 0.009 & 0.237 ± 0.025 & 0.058 ± 0.014 & - & - & - \\
BepiPred-3.0 & 0.276 ± 0.007 & 0.209 ± 0.011 & 0.194 ± 0.003 & 0.299 ± 0.009 & 0.087 ± 0.001 & 0.153 ± 0.005 \\
SEMA-1D 2.0 & 0.270 ± 0.005 & 0.213 ± 0.006 & 0.188 ± 0.002 & 0.327 ± 0.008 & 0.106 ± 0.007 & 0.193 ± 0.004 \\
DiscoTope-3.0 & 0.303 ± 0.007 & 0.231 ± 0.004 & 0.227 ± 0.007 & 0.396 ± 0.023 & 0.095 ± 0.007 & 0.307 ± 0.006 \\
BConformeR & \textbf{0.349 ± 0.004} & \textbf{0.262 ± 0.003} & \textbf{0.296 ± 0.003} & \textbf{0.405 ± 0.004} & \textbf{0.135 ± 0.014} & \textbf{0.350 ± 0.014} \\
\bottomrule
\end{tabular}
}
\caption{Performance comparison of different methods. F1-L and F1-D refer to the F1 scores calculated on linear and discontinuous epitopes, respectively. Standard deviation calculated over a range of thresholds.}
\label{table2}
\end{table*}

% %------------------------------------------------------------------------

\subsection{Comparison with SOTA Methods} 
\subsubsection{Benchmark evaluation}
BConformeR outperforms state-of-the-art baselines, achieving superior ROC-AUC and PR-AUC on Alphafold-predicted structures (Fig.~\ref{fig3}E,F) and higher PR-AUC on SARS-CoV-2 test set (Fig.~\ref{fig3}G). Notably, it consistently improves the prediction of both linear and discontinuous epitopes (Tab.~\ref{table2}), demonstrating robust identification of epitope regions and fragments.

% %------------------------------------------------------------------------

\begin{table}[!b]
\centering
\renewcommand{\arraystretch}{1}
\resizebox{0.85\textwidth}{!}{
\begin{tabular}{
    >{\centering\arraybackslash}p{2.25cm}|
    >{\centering\arraybackslash}p{0.95cm}
    >{\centering\arraybackslash}p{0.9cm}
    >{\centering\arraybackslash}p{2.0cm}|
    >{\centering\arraybackslash}p{0.95cm}
    >{\centering\arraybackslash}p{0.9cm}
    >{\centering\arraybackslash}p{2.2cm}|
    >{\centering\arraybackslash}p{1cm}|
    >{\centering\arraybackslash}p{1cm}
}
\toprule
\multirow{2}{*}{\textbf{Model}} 
  & \multicolumn{3}{c|}{\textbf{Linear}} 
  & \multicolumn{3}{c|}{\textbf{Discontinuous}} 
  & \multirow{2}{*}{\shortstack{\textbf{Params} \\ \textbf{(M)}}} 
  & \multirow{2}{*}{\shortstack{\textbf{MACs} \\ \textbf{(G)}}} \\
\cline{2-4} \cline{5-7}
& \textbf{Prec $\uparrow$} & \textbf{Rec $\uparrow$} & \textbf{F1 $\uparrow$} 
& \textbf{Prec $\uparrow$} & \textbf{Rec $\uparrow$} & \textbf{F1 $\uparrow$} 
&  &  \\
\midrule
BConformer-12 & 0.440 & 0.477 & 0.458 & 0.304 & 0.360 & 0.330 & 289.31 & 296.64    \\
BConformer-9  & 0.428 & 0.463 & 0.445(-2.8\%) & 0.297 & 0.322 & 0.309(-6.4\%) & 219.94 & 225.50   \\
BConformer-6  & 0.423 & 0.416 & 0.419(-8.5\%) & 0.249 & 0.296 & 0.270(-18.2\%) & 150.56 & 154.37    \\
ResNet1D-152   &  0.442  & 0.484 & 0.462(+0.9\%) & 0.255 & 0.291 & 0.272(-17.6\%) & 196.41 & 201.45 \\
ResNet1D-101    & 0.417 & 0.430 & 0.423(-7.6\%) & 0.242 & 0.238 & 0.240(-27.3\%) & 145.21 & 148.92 \\
ViT1D-12        & 0.429 & 0.435 & 0.432(-5.7\%) & 0.269 & 0.334 & 0.298(-9.7\%) & 230.05 & 234.22 \\
ViT1D-9        & 0.403 & 0.428 & 0.415(-9.4\%) & 0.276 & 0.311 & 0.292(-11.5\%) & 173.37 & 176.12 \\
\bottomrule
\addlinespace[2mm]
\end{tabular}
}
\caption{Ablation studies evaluated on the validation set. BConformer-12 corresponds to BConformeR, and BConformer-9/6 reduces one layer of FCU-based CNN and Transformer for each stage compared to BConformer-12/9. ResNet bottlenecks adopt the same CNN blocks as Conformer, and ViT follows the design of Transformer branch.}
\label{table3}
\end{table}

\subsection{Ablation}
\subsubsection{Overview} To further assess the contribution of each module to linear and conformational epitope prediction, we ablated the dual-branch Conformer into standalone ResNet and ViT variants. We first reduced the depth of BConformer-12 (BConformeR) to evaluate the effectiveness of the original architecture. Then, ResNet1D-152 and ResNet1D-101, as well as ViT1D-12 and ViT1D-9, were selected based on comparable parameters (Params) and multiply-accumulate operations (MACs) to BConformer-9 and BConformer-6 respectively. The F1 score was used as the primary evaluation metric for both linear and discontinuous epitope prediction.

% %------------------------------------------------------------------------

\begin{figure*}[t]
\centering
\includegraphics[width=1\textwidth]{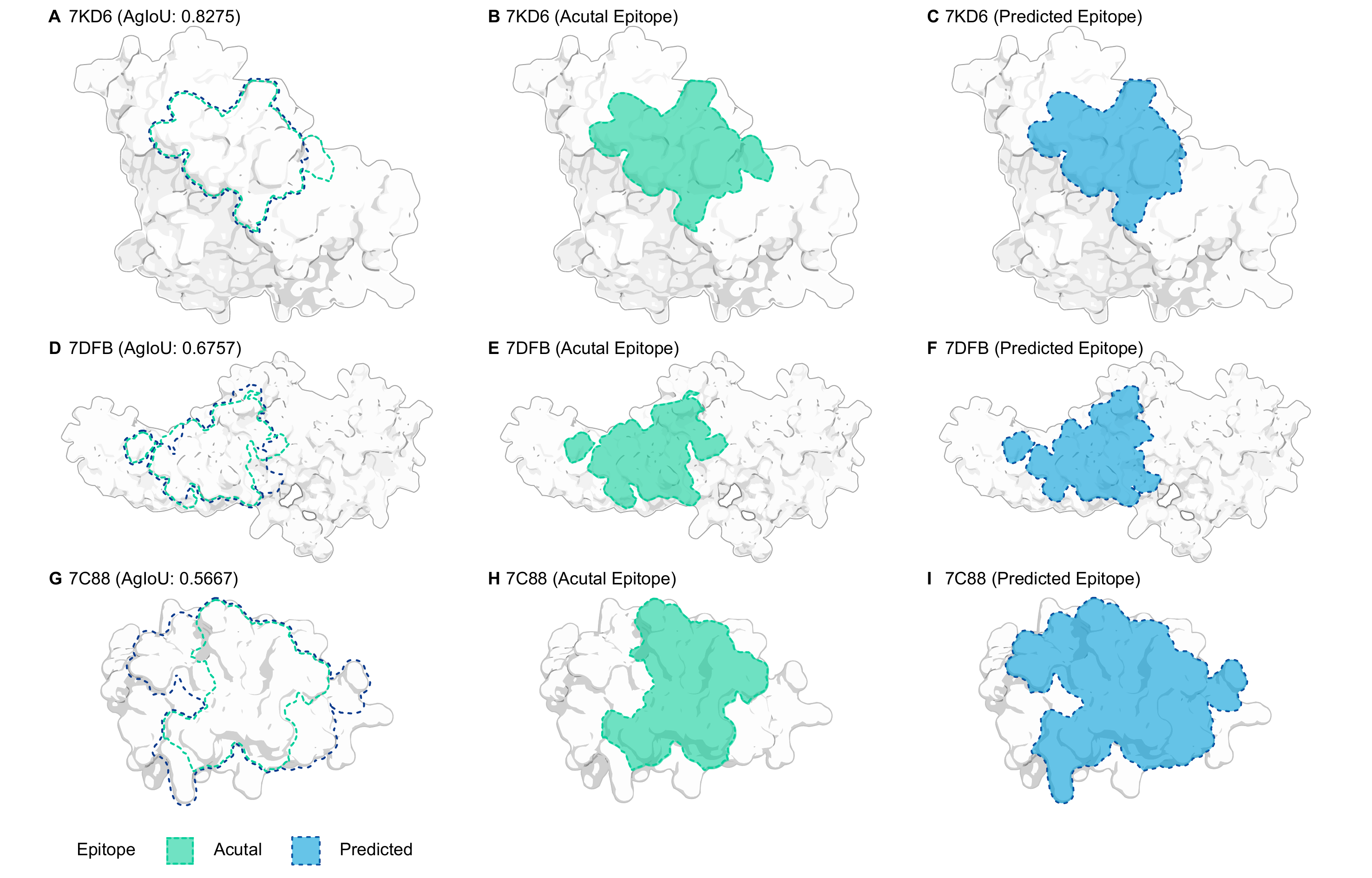} % Reduce the figure size so that it is slightly narrower than the column.
\caption{Structural visualization of epitopes predicted by BConformeR vs. actual epitopes. (A-C) Comparison of predicted and actual epitopes of 7KDB. Prediction is fully covered by the ground truth. (D-F) Comparison of predicted and actual epitopes of 7DFB. Prediction partially overlaps with the ground truth. (G-I) Comparison of predicted and actual epitopes of 7C88. Real epitopes are fully covered by the prediction.}
\label{fig4}
\end{figure*}

% %------------------------------------------------------------------------

% \subsubsection{Depth reduction} To evaluate the impact of model depth, we progressively reduced the number of layers in the Conformer architecture. As shown in Table~\ref{table3}, reducing depth to BConformer-9 leads to a moderate drop in performance, with linear and discontinuous F1 scores decreasing by 2.8\% (0.445) and 6.4\% (0.309) respectively. Further reduction to BConformer-6 causes a more significant decline, especially in discontinuous epitope prediction, where the F1 score drops by 18.2\% to 0.270. These demonstrate that our final model, BConformeR, achieves superior performance compared to shallower variants.

\subsubsection{Conformer vs.\ CNN}
Purely convolutional variants capture contiguous residue patterns efficiently and therefore achieve competitive scores on linear epitopes. However, their strictly local receptive field prevents them from modelling long‐range residue couplings that dominate conformational epitopes. As shown in Table~\ref{table3}, ResNet1D-152 attains the highest linear F1 score (0.462) among all models, confirming its effectiveness on local patterns. However, its discontinuous F1 drops to 0.272, lower than BConformer-9 (0.309) and ViT1D-12 (0.298), pointing to difficulty in capturing discontinuous dependencies.

By injecting a Transformer branch and exchanging information through FCUs, BConformer-9 achieves a noticeably higher discontinuous F1 than ResNet1D-152, while keeping a linear F1 of 0.445 with only a minor drop. This shows that substituting part of the convolutional layers with a Transformer modules improves discontinuous epitope prediction with little sacrifice in linear performance.

\subsubsection{Conformer vs.\ ViT}
Vision Transformer variants excel at aggregating sequence-wide information but lack explicit inductive bias for short-range motifs, often resulting in underfitting on linear stretches and noisy attention over densely packed regions. As shown in Table~\ref{table3},  ViT1D-9 achieves a discontinuous F1 of 0.292, surpassing ResNet1D-101 (0.240), ResNet1D-152 (0.272) and BConformer-6 (0.270). However, this comes at the cost of reduced performance on linear epitopes, where ViT1D-9 (0.415) trails behind ResNet1D-101 (0.423) and BConformer-6 (0.419).

Conformer mitigates this trade-off, as BConformer-9 outperforms ViT1D-12 on both linear and discontinuous F1 scores. Empirically, the hybrid design offers a more balanced solution for linear and discontinuous epitope prediction.

% %------------------------------------------------------------------------

\subsection{Case Studies}
\subsubsection{Predicted epitopes with high AgIoU} To visualize the prediction compared to the ground truth, we selected three antigen records from the test set with high AgIoU scores. Three overlap cases were shown in Figure~\ref{fig4}: (1) the prediction is fully covered by the real epitopes (Fig.~\ref{fig4}A-C), (2) the prediction partially overlaps with the real epitopes (Fig.~\ref{fig4}D-F), (3) the real epitopes are entirely encompassed by the prediction (Fig.~\ref{fig4}G-I).

% %------------------------------------------------------------------------
\section{Related Work}

Numerous antibody-agnostic epitope prediction methods have been proposed. SEPPA-3.0 was the epitope prediction tool specifically designed for glycoproteins, based on the logistic regression classifier \cite{SEPPA3.0}. BepiPred-3.0 used embeddings from pretrained protein language models ESM-2 as input to a feedforward neural network (FFNN) to predict both linear and conformational B-cell epitopes \cite{BepiPred3.0}. Epitope3D modeled antigen surfaces as residue-level graphs and extracted distance-based graph signatures to train classifiers for conformational B-cell epitope prediction, with AdaBoost selected as the backbone \cite{Epitope3D,AdaBoost}. ScanNet was an end-to-end geometric deep learning model that learned interpretable residue-level features directly from 3D structural neighborhoods \cite{ScanNet}. SEMA-1D 2.0 was constructed by adding a fully-connected linear layer on an ensemble of five ESM-2 models, while SEMA-3D 2.0 was developed upon an ensemble of five pre-trained SaProt models \cite{SEMA2.0,SaProt}. DiscoTope-3.0 utilized inverse folding representations from ESM-IF1 and was trained on both predicted and solved structures using a positive-unlabelled ensemble strategy, enabling structure-based B-cell epitope prediction at scale \cite{DiscoTope3.0,ESM-IF1}.

% %------------------------------------------------------------------------
\section{Conclusion}

We present \textbf{BConformeR}, a dual–branch Conformer model that integrates convolutional and self–attention mechanisms for residue–level prediction of both linear and conformational B-cell epitopes from antigen sequences. Two contributions are worth noting: 
% (i) a bidirectional \emph{feature–coupling} strategy that lets local and global representations refine each other;  
(i) Adaptive \emph{feature coupling and logit fusion} that balance information from the CNN and Transformer branches, yielding a well-calibrated score for each residue;
(ii) a detailed \emph{performance analysis on both linear and discontinuous epitopes}, which also validates the complementary effectiveness of both branches.
%(ii) \emph{AgIoU}, an antigen–level overlap metric that provides an intuitive, single–value summary of epitope prediction quality.  
Benchmark evaluation demonstrates that our model consistently outperforms strong sequence- and structure-based baselines, particularly on the more difficult task of predicting conformational epitopes. The anticipated beneficiaries include rational vaccine design, diagnostic reagent development and antibody engineering pipelines.

\appendix
\section*{Acknowledgments}
The computation resources in this study was sponsored by BayVax Biotech Limited. Y.-F. H. thanks support from the Major Program of the National Natural Science Foundation of China (\texttt{92369201}). The research was also supported by the National Key Research and Development Program of China (\texttt{2021YFA0910700}), the Health and Medical Research Fund, the Food and Health Bureau, The Government of the Hong Kong Special Administrative Region (\texttt{COVID1903010},  \texttt{T-11-709/21-N}) to J.-D.H. J.-D.H. also thanks the L \& T Charitable Foundation, the Program for Guangdong Introducing Innovative and Entrepreneurial Teams (\texttt{2019BT02Y198}) and Shenzhen Key Laboratory for Cancer Metastasis and Personalized Therapy (\texttt{ZDSYS20210623091811035}) for their support. 
%Bibliography
\bibliographystyle{unsrt}  
\bibliography{references} 

@article{koz:18,
  title={Computational B-cell epitope identification and production of neutralizing murine antibodies against Atroxlysin-I},
  author={Kozlova, Edgar Ernesto Gonzalez and Cerf, Lo{\"\i}c and Schneider, Francisco Santos and Viart, Benjamin Thomas and NGuyen, Christophe and Steiner, Bethina Trevisol and de Almeida Lima, Sabrina and Molina, Franck and Duarte, Clara Guerra and Felicori, Liza and others},
  journal={Scientific reports},
  volume={8},
  number={1},
  pages={14904},
  year={2018},
  publisher={Nature Publishing Group UK London}
}

@article{ji:22,
  title={Identification of linear B cell epitopes on CD2V protein of African swine fever virus by monoclonal antibodies},
  author={Jia, Rui and Zhang, Gaiping and Bai, Yilin and Liu, Hongliang and Chen, Yumei and Ding, Peiyang and Zhou, Jingming and Feng, Hua and Li, Mingyang and Tian, Yuanyuan and others},
  journal={Microbiology Spectrum},
  volume={10},
  number={2},
  pages={e01052--21},
  year={2022},
  publisher={American Society for Microbiology 1752 N St., NW, Washington, DC}
}

@article{va:22,
  title={Evaluation from a B-cell epitope-based chimeric protein for the serodiagnosis of tegumentary and visceral leishmaniasis},
  author={Vale, Danniele L and Machado, Amanda S and Ramos, Fernanda F and Lage, Daniela P and Freitas, Camila S and de Oliveira, Daysiane and Galvani, Nathalia C and Luiz, Gabriel P and Fagundes, Mirian I and Fernandes, Bruna B and others},
  journal={Microbial Pathogenesis},
  volume={167},
  pages={105562},
  year={2022},
  publisher={Elsevier}
}

@article{ta:24,
  title={Epitope landscape in autoimmune neurological disease and beyond},
  author={Talucci, Ivan and Maric, Hans M},
  journal={Trends in pharmacological sciences},
  volume={45},
  number={9},
  pages={768--780},
  year={2024},
  publisher={Elsevier}
}

@article{dv:24,
  title={Conversion of vaccines from low to high immunogenicity by antibodies with epitope complementarity},
  author={Dvorscek, Alexandra R and McKenzie, Craig I and St{\"a}heli, Vera C and Ding, Zhoujie and White, Jacqueline and Fabb, Stewart A and Lim, Leonard and O’Donnell, Kristy and Pitt, Catherine and Christ, Daniel and others},
  journal={Immunity},
  volume={57},
  number={10},
  pages={2433--2452},
  year={2024},
  publisher={Elsevier}
}

@article{ba:25,
  title={Computational epitope-based vaccine design with bioinformatics approach; a review},
  author={Basmenj, Esmaeil Roohparvar and Pajhouh, Susan Radman and Fallah, Afsane Ebrahimi and Rahimi, Elmira and Atighy, Hossein and Ghiabi, Shadan and Ghiabi, Shamim and others},
  journal={Heliyon},
  volume={11},
  number={1},
  year={2025},
  publisher={Elsevier}
}

@article{crys:21,
  title={Advances and challenges in time-resolved macromolecular crystallography},
  author={Br{\"a}nd{\'e}n, Gisela and Neutze, Richard},
  journal={Science},
  volume={373},
  number={6558},
  pages={eaba0954},
  year={2021},
  publisher={American Association for the Advancement of Science}
}

@article{cryoEM:25,
  title={Advances in cryo-electron microscopy (cryoEM) for structure-based drug discovery},
  author={Rubach, Pawel and Majorek, Karolina A and Gucwa, Michal and Murzyn, Krzysztof and Wlodawer, Alexander and Minor, Wladek},
  journal={Expert Opinion on Drug Discovery},
  number={just-accepted},
  year={2025},
  publisher={Taylor \& Francis}
}

@article{pd:22,
  title={Advances in antibody phage display technology},
  author={Ledsgaard, Line and Ljungars, Anne and Rimbault, Charlotte and S{\o}rensen, Christoffer V and Tulika, Tulika and Wade, Jack and Wouters, Yessica and McCafferty, John and Laustsen, Andreas H},
  journal={Drug Discovery Today},
  volume={27},
  number={8},
  pages={2151--2169},
  year={2022},
  publisher={Elsevier}
}

@article{BepiPred1.0,
  title={Improved method for predicting linear B-cell epitopes},
  author={Larsen, Jens Erik Pontoppidan and Lund, Ole and Nielsen, Morten},
  journal={Immunome research},
  volume={2},
  pages={1--7},
  year={2006},
  publisher={Springer}
}

@article{BepiPred2.0,
  title={BepiPred-2.0: improving sequence-based B-cell epitope prediction using conformational epitopes},
  author={Jespersen, Martin Closter and Peters, Bjoern and Nielsen, Morten and Marcatili, Paolo},
  journal={Nucleic acids research},
  volume={45},
  number={W1},
  pages={W24--W29},
  year={2017},
  publisher={Oxford University Press}
}

@article{BepiPred3.0,
  title={BepiPred-3.0: Improved B-cell epitope prediction using protein language models},
  author={Clifford, Joakim N{\o}ddeskov and H{\o}ie, Magnus Haraldson and Deleuran, Sebastian and Peters, Bjoern and Nielsen, Morten and Marcatili, Paolo},
  journal={Protein Science},
  volume={31},
  number={12},
  pages={e4497},
  year={2022},
  publisher={Wiley Online Library}
}

@article{DiscoTope2.0,
  title={Reliable B cell epitope predictions: impacts of method development and improved benchmarking},
  author={Kringelum, Jens Vindahl and Lundegaard, Claus and Lund, Ole and Nielsen, Morten},
  journal={PLoS computational biology},
  volume={8},
  number={12},
  pages={e1002829},
  year={2012},
  publisher={Public Library of Science San Francisco, USA}
}

@article{DiscoTope3.0,
  title={DiscoTope-3.0: improved B-cell epitope prediction using inverse folding latent representations},
  author={H{\o}ie, Magnus Haraldson and Gade, Frederik Steensgaard and Johansen, Julie Maria and W{\"u}rtzen, Charlotte and Winther, Ole and Nielsen, Morten and Marcatili, Paolo},
  journal={Frontiers in immunology},
  volume={15},
  pages={1322712},
  year={2024},
  publisher={Frontiers Media SA}
}

@article{SEPPA3.0,
  title={SEPPA 3.0—enhanced spatial epitope prediction enabling glycoprotein antigens},
  author={Zhou, Chen and Chen, Zikun and Zhang, Lu and Yan, Deyu and Mao, Tiantian and Tang, Kailin and Qiu, Tianyi and Cao, Zhiwei},
  journal={Nucleic acids research},
  volume={47},
  number={W1},
  pages={W388--W394},
  year={2019},
  publisher={Oxford University Press}
}

@article{SEMA,
  title={SEMA: Antigen B-cell conformational epitope prediction using deep transfer learning},
  author={Shashkova, Tatiana I and Umerenkov, Dmitriy and Salnikov, Mikhail and Strashnov, Pavel V and Konstantinova, Alina V and Lebed, Ivan and Shcherbinin, Dmitriy N and Asatryan, Marina N and Kardymon, Olga L and Ivanisenko, Nikita V},
  journal={Frontiers in immunology},
  volume={13},
  pages={960985},
  year={2022},
  publisher={Frontiers Media SA}
}

@article{SEMA2.0,
  title={SEMA 2.0: web-platform for B-cell conformational epitopes prediction using artificial intelligence},
  author={Ivanisenko, Nikita V and Shashkova, Tatiana I and Shevtsov, Andrey and Sindeeva, Maria and Umerenkov, Dmitriy and Kardymon, Olga},
  journal={Nucleic Acids Research},
  volume={52},
  number={W1},
  pages={W533--W539},
  year={2024},
  publisher={Oxford University Press}
}

@article{Epitope3D,
  title={epitope3D: a machine learning method for conformational B-cell epitope prediction},
  author={da Silva, Bruna Moreira and Myung, YooChan and Ascher, David B and Pires, Douglas EV},
  journal={Briefings in Bioinformatics},
  volume={23},
  number={1},
  year={2022},
  publisher={Oxford Academic}
}

@article{ScanNet,
  title={ScanNet: an interpretable geometric deep learning model for structure-based protein binding site prediction},
  author={Tubiana, J{\'e}r{\^o}me and Schneidman-Duhovny, Dina and Wolfson, Haim J},
  journal={Nature Methods},
  volume={19},
  number={6},
  pages={730--739},
  year={2022},
  publisher={Nature Publishing Group US New York}
}

@article{AdaBoost,
  title={Advance and prospects of AdaBoost algorithm},
  author={Ying, Cao and Qi-Guang, Miao and Jia-Chen, Liu and Lin, Gao},
  journal={Acta Automatica Sinica},
  volume={39},
  number={6},
  pages={745--758},
  year={2013},
  publisher={Elsevier}
}

@article{ESM-2pre,
  title={Language models of protein sequences at the scale of evolution enable accurate structure prediction},
  author={Lin, Zeming and Akin, Halil and Rao, Roshan and Hie, Brian and Zhu, Zhongkai and Lu, Wenting and dos Santos Costa, Allan and Fazel-Zarandi, Maryam and Sercu, Tom and Candido, Sal and others},
  journal={BioRxiv},
  volume={2022},
  pages={500902},
  year={2022}
}

@inproceedings{ESM-IF1,
  title={Learning inverse folding from millions of predicted structures},
  author={Hsu, Chloe and Verkuil, Robert and Liu, Jason and Lin, Zeming and Hie, Brian and Sercu, Tom and Lerer, Adam and Rives, Alexander},
  booktitle={International conference on machine learning},
  pages={8946--8970},
  year={2022},
  organization={PMLR}
}

@article{ESM-2,
  title={Evolutionary-scale prediction of atomic-level protein structure with a language model},
  author={Lin, Zeming and Akin, Halil and Rao, Roshan and Hie, Brian and Zhu, Zhongkai and Lu, Wenting and Smetanin, Nikita and Verkuil, Robert and Kabeli, Ori and Shmueli, Yaniv and others},
  journal={Science},
  volume={379},
  number={6637},
  pages={1123--1130},
  year={2023},
  publisher={American Association for the Advancement of Science}
}

@article{OverviewEp,
  title={An Overview of Epitopes and Methods for Antibody Epitope Mapping},
  author={Nilvebrant, Johan and Berndt Thal{\'e}n, Niklas and Rockberg, Johan and Malm, Magdalena},
  journal={Epitope Mapping Protocols},
  pages={1--11},
  year={2025},
  publisher={Springer}
}

@article{AbEpitope,
  title={AbEpiTope-1.0: Improved antibody target prediction by use of AlphaFold and inverse folding},
  author={Clifford, Joakim N{\o}ddeskov and Richardson, Eve and Peters, Bjoern and Nielsen, Morten},
  journal={Science Advances},
  volume={11},
  number={24},
  pages={eadu1823},
  year={2025},
  publisher={American Association for the Advancement of Science}
}

@article{Conformer20,
  title={Conformer: Convolution-augmented transformer for speech recognition},
  author={Gulati, Anmol and Qin, James and Chiu, Chung-Cheng and Parmar, Niki and Zhang, Yu and Yu, Jiahui and Han, Wei and Wang, Shibo and Zhang, Zhengdong and Wu, Yonghui and others},
  journal={arXiv preprint arXiv:2005.08100},
  year={2020}
}

@inproceedings{Conformer21,
  title={Conformer: Local features coupling global representations for visual recognition},
  author={Peng, Zhiliang and Huang, Wei and Gu, Shanzhi and Xie, Lingxi and Wang, Yaowei and Jiao, Jianbin and Ye, Qixiang},
  booktitle={Proceedings of the IEEE/CVF international conference on computer vision},
  pages={367--376},
  year={2021}
}

@article{SaProt,
  title={Saprot: Protein language modeling with structure-aware vocabulary},
  author={Su, Jin and Han, Chenchen and Zhou, Yuyang and Shan, Junjie and Zhou, Xibin and Yuan, Fajie},
  journal={BioRxiv},
  pages={2023--10},
  year={2023},
  publisher={Cold Spring Harbor Laboratory}
}

@article{epitope_id,
  title={Does difference exist between epitope and non-epitope residues?},
  author={Sun, Jing and Xu, Tianlei and Wang, Shuning and Li, Guoqing and Wu, Di and Cao, Zhiwei},
  journal={Immunome research},
  volume={7},
  number={3},
  pages={1},
  year={2011},
  publisher={International Immunomics Society}
}

@article{epitope_cl,
  title={B-cell epitopes: Discontinuity and conformational analysis},
  author={Ferdous, Saba and Kelm, Sebastian and Baker, Terry S and Shi, Jiye and Martin, Andrew CR},
  journal={Molecular immunology},
  volume={114},
  pages={643--650},
  year={2019},
  publisher={Elsevier}
}

@inproceedings{ResNet,
  title={Deep residual learning for image recognition},
  author={He, Kaiming and Zhang, Xiangyu and Ren, Shaoqing and Sun, Jian},
  booktitle={Proceedings of the IEEE conference on computer vision and pattern recognition},
  pages={770--778},
  year={2016}
}

@article{ViT,
  title={An image is worth 16x16 words: Transformers for image recognition at scale},
  author={Dosovitskiy, Alexey and Beyer, Lucas and Kolesnikov, Alexander and Weissenborn, Dirk and Zhai, Xiaohua and Unterthiner, Thomas and Dehghani, Mostafa and Minderer, Matthias and Heigold, Georg and Gelly, Sylvain and others},
  journal={arXiv preprint arXiv:2010.11929},
  year={2020}
}

@article{Attention,
  title={Attention is all you need},
  author={Vaswani, Ashish and Shazeer, Noam and Parmar, Niki and Uszkoreit, Jakob and Jones, Llion and Gomez, Aidan N and Kaiser, {\L}ukasz and Polosukhin, Illia},
  journal={Advances in neural information processing systems},
  volume={30},
  year={2017}
}

@article{AACDB_data,
  title={A comprehensive antigen-antibody complex database unlocking insights into interaction interface},
  author={Zhou, Yuwei and Liu, Wenwen and Huang, Ziru and Gou, Yushu and Liu, Siqi and Jiang, Lixu and Yang, Yue and Huang, Jian},
  journal={eLife},
  volume={14},
  pages={RP104934},
  year={2025},
  publisher={eLife Sciences Publications Limited}
}

@article{CoV-AbDab,
  title={CoV-AbDab: the coronavirus antibody database},
  author={Raybould, Matthew IJ and Kovaltsuk, Aleksandr and Marks, Claire and Deane, Charlotte M},
  journal={Bioinformatics},
  volume={37},
  number={5},
  pages={734--735},
  year={2021},
  publisher={Oxford University Press}
}

@article{GAM,
  title={pygam: Generalized additive models in python},
  author={Serv{\'e}n, Daniel and Brummitt, Charlie},
  journal={Zenodo},
  year={2018}
}

@article{AFMultimer,
  title={ColabFold: making protein folding accessible to all},
  author={Mirdita, Milot and Sch{\"u}tze, Konstantin and Moriwaki, Yoshitaka and Heo, Lim and Ovchinnikov, Sergey and Steinegger, Martin},
  journal={Nature methods},
  volume={19},
  number={6},
  pages={679--682},
  year={2022},
  publisher={Nature Publishing Group US New York}
}

\clearpage

% Below is the appendix
% Below is the appendix
% Below is the appendix
% Below is the appendix
% Below is the appendix

% %------------------------------------------------------------------------

\appendix
\section*{Appendix}

\begin{figure}[H]
\centering
\includegraphics[width=0.7\linewidth]{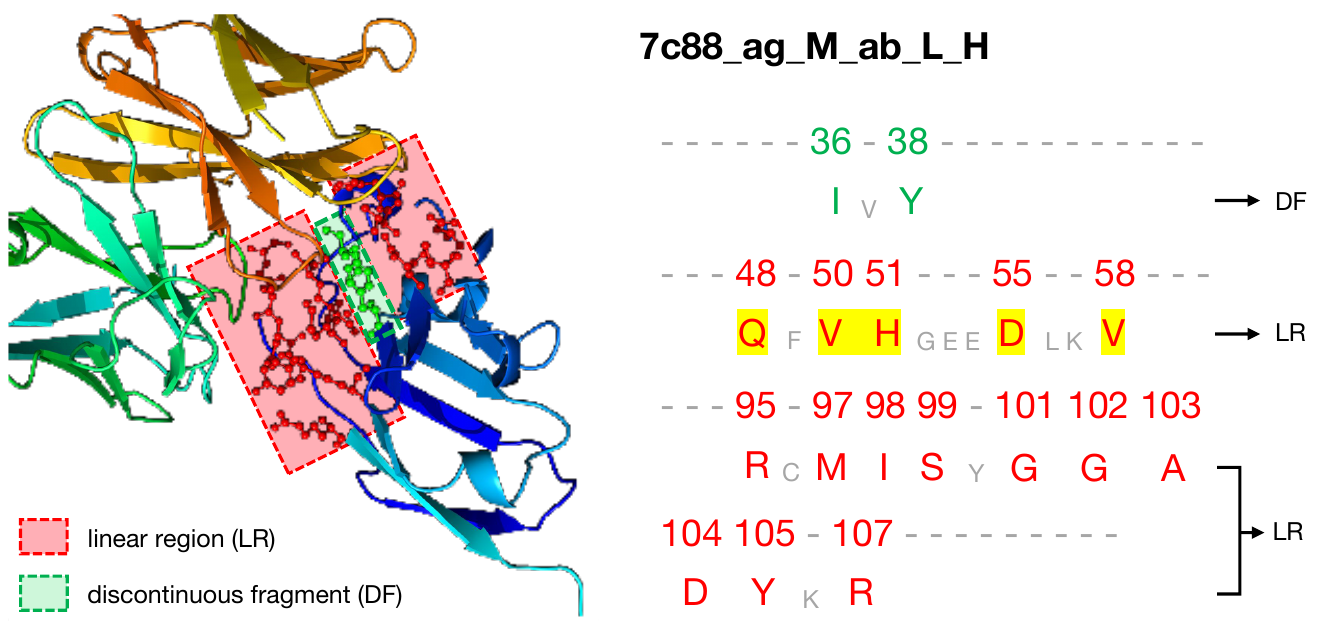} % Reduce the figure size so that it is slightly narrower than the column.
\caption{Linear regions and discontinuous fragments. (Left) The blue structure represents antigen 7c88, and the light green and light orange correspond to part of the antibody light chain and heavy chain, respectively; Linear regions are marked with red rectangles, and discontinuous fragments with green. (Right) Residues highlighted with numbers indicate epitopes.}
\label{ap_fig2}
\end{figure}

% %------------------------------------------------------------------------
\section{Linear Regions and Discontinuous Fragments}
\label{ap_LRDF}
Regions are defined as linear stretches of antigen sequence having at least three antibody-contacting residues. Gaps between contacting residues are allowed, and a gap size of up to three non-contacting residues is chosen on the basis of the structure of $\alpha$-helices, which enables residues on the same helical face to be included. Fragments, which do not contain enough antibody-contacting residues to qualify as a region (less than three), are considered discontinuous in our study (Fig.~\ref{ap_fig2}).

% %------------------------------------------------------------------------

\section{Performance Metrics}
\label{ap_metrics}
The performance metrics used in this study include Precision (Prec), Recall (Rec), True Positive Rate (TPR), False Positive Rate (FPR), F1-score, Antigen-level intersection over union (AgIoU), Matthews correlation coefficient (MCC), whose definitions are given as follows: 
\begin{equation*}
\begin{aligned}
&\text{Prec} = \frac{\text{TP}}{\text{TP} + \text{FP}}, \\[3pt]
&\text{Rec} = \frac{\text{TP}}{\text{TP} + \text{FN}}, \\[3pt]
&\text{TPR} = \frac{\text{TP}}{\text{TP} + \text{FN}}, \\[3pt]
&\text{FPR} = \frac{\text{FP}}{\text{FP} + \text{TN}}, \\[3pt]
&\text{F1} = \frac{2\text{TP}}{2\text{TP} + \text{FP} + \text{FN}}, \\[3pt]
&\text{AgIoU} = \frac{\text{TP}}{\text{TP} + \text{FP} + \text{FN}}, \\[3pt]
&\text{MCC} = \frac{\text{TP} \cdot \text{TN} - \text{FP} \cdot \text{FN}}{\sqrt{(\text{TP}+\text{FP})(\text{TP}+\text{FN})(\text{TN}+\text{FP})(\text{TN}+\text{FN})}},
\end{aligned}
\end{equation*}
where \(\text{TP}\), \(\text{TN}\), \(\text{FP}\) and \(\text{FN}\) denote the number of true positives, true negatives, false positives and false negatives, respectively. The relationships between some of these metrics are given as:
\begin{equation*}
\begin{aligned}
&\text{F1} = \frac{2 \cdot \text{Prec} \cdot \text{Rec}}
{\text{Prec} + \text{Rec}} = \frac{2 \cdot \text{AgIoU}}{1 + \text{AgIoU}}, \\[3pt]
&\text{AgIoU} = \frac{2 \cdot \text{F1}}{2 - \text{F1}}. \\[3pt]
\end{aligned}
\end{equation*}

% %------------------------------------------------------------------------

\section{Ablation}
\label{ap_ablation}
\subsection{ViT1D} 

\begin{table}[H]
\centering
\renewcommand{\arraystretch}{1}
\begin{tabular}{l@{\hskip 10pt}c@{\hskip 11pt}c@{\hskip 11pt}c@{\hskip 11pt}c@{\hskip 9pt}c@{\hskip 9pt}c@{\hskip 9pt}c}
\toprule
\textbf{Model} & \textbf{Embed Dim} & \textbf{Patch Size} & \textbf{Depth} & \textbf{Heads} & \textbf{MLP Ratio} \\
\midrule
ViT1D-9  & 1536 & 1 & 9  & 12 & 2.0 \\
ViT1D-12 & 1536 & 1 & 12 & 12 & 2.0 \\
\bottomrule
\addlinespace[2mm]
\end{tabular}
\caption{ViT variants used in ablation study.}
\label{ap_vit1d}
\end{table}

We conducted ablation studies on two variants of the ViT1D model differing in depth: ViT1D-9 and ViT1D-12. As shown in Table~\ref{ap_vit1d}, both models share the same embedding dimension (1536), input sequence length (1024), patch size (1), number of attention heads (12), and MLP expansion ratio (2.0). The only difference is the number of transformer encoder layers.

\subsection{ResNet1D}

\renewcommand{\arraystretch}{1.05}
\begin{table}[H]
\centering
\begin{tabular}{l@{\hskip 6pt}c@{\hskip 9pt}c@{\hskip 9pt}c@{\hskip 9pt}c}
\toprule
\textbf{Stage} & \textbf{Output Channels} & \textbf{Kernel Size} & \textbf{ResNet1D-101} & \textbf{ResNet1D-152} \\
\midrule
input & 1152 & - & $\checkmark$ & $\checkmark$ \\
stem & 288 & 7 & $\checkmark$ & $\checkmark$ \\
layer1 & 576 (144 $\times$ 4) & 3 & 3 & 3 \\
layer2 & 1152 (288 $\times$ 4) & 3 & 4 & 8 \\
layer3 & 2304 (576 $\times$ 4) & 3 & 23 & 36 \\
layer4 & 4608 (1152 $\times$ 4) & 3 & 3 & 3 \\
classifier & 2 & 1 & $\checkmark$ & $\checkmark$ \\
\bottomrule
\addlinespace[2mm]
\end{tabular}
\caption{ResNet architectures used in ablation study.}
\label{ap_resnet1d}
\end{table}

We employed two variants of the ResNet1D backbone in the ablation studies: ResNet1D-101 and ResNet1D-152. As shown in Table~\ref{ap_resnet1d}, both models accept input feature maps with 1152 channels and begin with an initial convolutional layer producing 320 channels. The networks consist of four residual stages of bottlenecks with the expansion factor of 4, each stage starting by downsampling the channel dimension to half. The difference lies in the number of bottleneck blocks per stage. ResNet1D-101 contains [3, 4, 23, 3] blocks across the four layers, while ResNet1D-152 is deeper with [3, 8, 36, 3] blocks. The final output is projected to the target classes through a 1$\times$1 convolutional layer.
% %------------------------------------------------------------------------
.

\end{document}